\documentclass[12pt]{article}
\usepackage[utf8]{inputenc}
\usepackage[english]{babel}
\usepackage{amsmath, amssymb, graphicx}
\renewcommand{\theequation}{\arabic{section}.\arabic{equation}}
\usepackage{hyperref}
\usepackage{braket}
\usepackage{color}
\usepackage{xcolor}

\newcommand{\be}{\begin{equation}}
\newcommand{\ee}{\end{equation}}
\newcommand{\bea}{\begin{eqnarray}}
\newcommand{\eea}{\end{eqnarray}}

\newcommand{\fnm}{\footnotemark}
\newcommand{\fnt}{\footnotetext}

\begin{document}

\begin{center}
 
 \large \bf
Heavy quarks in rotating plasma via holography
\end{center}

\vspace{15pt}

\begin{center}

 \normalsize\bf
         Anastasia A. Golubtsova\fnm[1]\fnt[1]{golubtsova@theor.jinr.ru}$^{, a,b}$,
         Eric Gourgoulhon\fnm[2]\fnt[2]{eric.gourgoulhon@obspm.fr}$^{, c}$,
        Marina K. Usova\fnm[3]\fnt[3]{umk.16@uni-dubna.ru}$^{,a, b}$

 \vspace{7pt}

 \it (a) \ \ \ \ Bogoliubov Laboratory of Theoretical Physics, JINR,\\
Joliot-Curie str. 6,  Dubna, 141980  Russia  \\

(b) \ Dubna State University, Universitetskaya str. 19, \\
   Dubna, 141980, Russia
   
(c) \,Laboratoire Univers et Th\'{e}ories, Observatoire de Paris, CNRS,
Universit\'{e} PSL, Universit\'{e} de Paris,
5 place Jules Janssen, 92190 Meudon, France

 \end{center}
 \vspace{15pt}

\begin{abstract}
We continue to explore the dynamics of a heavy quark in a strongly coupled rotating quark-gluon plasma within the holographic framework. 
We use the 5d Kerr-AdS black hole with two arbitrary rotating parameters to describe the rotating quark-gluon plasma. 
We present the relations for the drag force acting on the heavy quark both in the rotating- and static-at-infinity frames for arbitrary choice of the rotational parameters.
 We also calculate  the thermal mass of a static quark  in different frames and show that at high-$T$ limit the dependence on rotation is suppressed.

\end{abstract}
 \newpage
\tableofcontents
\section{Introduction}

Studies of black holes in anti-de Sitter (AdS) spacetimes via holography are important since they help to understand features of dual field theories at strong coupling that are inaccessible by perturbative calculations \cite{CSLMRW}-\cite{AREF4}. Particularly, the holographic duality provides a link between properties of asymptotically AdS black holes  and observables relevant to physics of the quark-gluon plasma (QGP). The QGP represents a hot strongly coupled fluid in the phase of deconfinement and, moreover, at high temperature,
the lattice results show near conformal behaviour, so the holographic approach seems to be viable for QGP. 

Recently there has been some significant interest in the description and study of a rotating QGP formed in heavy-ion collisions.
 It is known that  in non-central heavy-ion collisions there is a nonzero total angular momentum that is related to colliding nuclei.
The major part of this angular momentum is taken away by the spectator nucleons; however some amount of angular momentum remains in the QGP and is conserved in time \cite{QGPR3}-\cite{QGPR2}.  The measurements of the $\Lambda$, $\bar{\Lambda}$ hyperon polarization by STAR collaboration at RHIC predict the angular velocity $\Omega \sim 6 \pm 1$ MeV \cite{Abelev:2007zk}-\cite{STAR:2017ckg}.  Note that the value of $\Omega$ obtained in hydrodynamic simulations is even greater: $\Omega \sim 20-40$ MeV \cite{QGPR2}.
 So it is of interest to trace how the fast rotation leads to different phenomena in QGP.
 
In the holographic context, the rotating QGP can be described by virtue of a 5d  asymptotically AdS black hole with a non-zero angular momentum \cite{2010.14478, BLLM}-\cite{2003.03765}. This class of black holes includes  the 5d Kerr-AdS solutions constructed in \cite{HHT}.  It is interesting to note that, comparing to the 4-dimensional Kerr-AdS black holes \cite{Carter}, the five dimensional ones have two rotational parameters $a$ and $b$, i.e. the angular momentum consists of two parts that are conserved independently. The 5-dimensional Kerr-AdS black holes  are dual to the thermal $\mathcal{N}=4$ SYM at strong coupling  on the cylinder $\mathbb{R}\times \mathbb{S}^{3}$ \cite{HHT, HReal}.   We note that the Kerr-$AdS_{5}$ black hole can be described in rotating-  (Boyer-Lindquist coordinates) or static-at-the-boundary (asymptotically AdS coordinates) frames. The disadvantage of BL coordinates comparing to  AdS ones is an absence of $S^3$-symmetry near the boundary of the Kerr-$AdS_{5}$ spacetime. However, for small values of rotational parameters the boundary in BL coordinates has a similar form to the boundary of the  Kerr-AdS spacetime in the static-at-the-boundary frame. The case where both rotational parameters are non-zero can be associated with a gyroscopic motion in the dual field theory, while a single non-zero rotational parameter corresponds to a rotation around some axis.
 The Kerr-AdS black holes possess the Hawking-Page transition, which is associated to a confinement-deconfinement phase transition in the dual theory. The rotation of the Kerr-AdS black hole affects the phase transition. Namely, the temperature of the phase transition decreases with the rotation, as shown using the holographic approach in the recent work \cite{2010.14478}. The similar influence of the rotation on the Hawking-Page transition was found in \cite{AGG2020}, where the behaviour of the free energy and the entropy was studied.  Moreover, in the latter work it was shown that  the phase transition takes place only for relatively small parameters and disappears if at least one of them is greater than some admissible value ($a,b \in[0; 0.25]$ for $\ell =1$). These values of the rotational parameters exactly correspond to the angular velocity obtained in the hydrodynamical simulations \cite{QGPR2}. 
It worth to be noted that in a number of the effective models  of rotating QGP \cite{Ebihara:2016fwa,Chernodub:2016kxh,Chernodub:2020qah, Fujimoto:2021xix} it was found that the rotation decreases the deconfinement temperature. 
At the same time, recent studies \cite{Braguta, Braguta2}
using lattice QCD calculations have shown that the phase transition temperature $T_{c}$ increases with $\Omega$.

Heavy quarks represent suitable probes of the strongly coupled QGP.  Transport properties of QGP can be understood by calculating observables related to heavy quark dynamics in the medium. When moving through the QGP, an energetic quark looses its energy fast because of strong coupling. Energy losses are observed through jet quenching at experiments on RHIC.  For heavy quarks, which have a relative large mass,  energy losses can be studied more clearly, since low-energy strong-interaction effects are masked by the quark mass.

Using a perturbative QCD approach, the energy of  the heavy quark momentum diffusion coefficient was calculated earlier in \cite{AMY}-\cite{CHM2}.  
Another possibility to study jet quenching at strong coupling is the lattice-QCD  calculations.  
The heavy quark diffusion coefficient and the jet quenching coefficient, related with the averaged transverse momentum broadening squared per unit mean free path  \cite{Wang:1992qdg,Baier:1996sk}, using the lattice approach were discussed in \cite{Laine:2009dd}-\cite{Panero:2013pla}.

 In the holographic framework the energy loss of the heavy quark can be studied by at least two ways.  
 The first one is based on the relation of a jet-quenching coefficient $\hat{q}$ to light-like Wilson loops and comes down to the calculation of holographic Wilson loops \cite{HoloJet}. 
 
  The second way is to calculate the drag force acting on a heavy quark in the thermal medium. In this case the heavy quark is associated  with an endpoint of an open string stretched in the black hole background from the boundary of the spacetime down to the horizon \cite{GubserDF, HKKKY}. Within the second approach in the ordinary and charged AdS black branes the energy loss was broadly studied in \cite{0605235}-\cite{DF-aniz-DG}.  In the framework of the fluid/gravity correspondence the drag force was studied in \cite{Rajagopal:2015roa, Reiten:2019fta}.

In works \cite{NAS,AGG2020} the holographic drag force  in the rotating QGP  was calculated using 4d  and 5d Kerr-AdS black holes, respectively. In the paper \cite{AGG2020} the drag force  was found for the cases of the 5d Kerr-AdS black hole with one non-zero and two equal rotational parameters.
It was shown in \cite{AGG2020} that if the 5d black hole has just one rotational parameter the relation for the drag force is in agreement with the result of \cite{NAS};
at the same time it was obtained that the drag force vanishes for the case of two equal rotational parameters.

The purpose of this paper is to present generic relations for the drag force acting on a heavy quark in a rotating quark-gluon plasma within the holographic correspondence for two arbitrary rotating parameters. For our analysis we consider dynamics of an open string, with one endpoint attached at the spacetime boundary. The endpoint is fixed, so in the Boyer-Lindquist coordinates, this corresponds to a frame rotating-at-infinity, the endpoint rotates together with the quark-gluon plasma, while in the AdS coordinates, which are related to a  static-at-infinity frame, the endpoint does not rotate, however  the ansatz for the dynamical variables can be interpreted as a slowly motion of the string.
According to the holographic dictionary, the conjugate momenta is related to a drag force and the total energy of the string corresponds to a thermal mass of a quark at rest.

We also  consider an open string with both endpoints located on the boundary of the Kerr-${\rm AdS}_{5}$ spacetime and hanging down to the black hole horizon in the static-at-infinity frame.
This string configuration can be associated with a  light-like Wilson loop, which allows to  estimate the jet-quenching parameter for a heavy quark in the rotating QGP. 

The paper is organized as follows.
First, in Section 2 we give a brief overview of an open string dynamics in the generic 5-dimensional gravitational background, we write down relations for the conjugate momenta and the total energy carried by the string. We also present the 5d Kerr-AdS black holes with two non-zero rotational parameters in Boyer-Lindquist coordinates (rotating-at-infinity), as well as in the so-called AdS coordinates (static at the boundary). In Section 3 we consider a straight static string, which holographically describes a static quark,  and find the string energy, which is interpreted as a thermal static mass of the quark. We calculate the string energy for both frames  of and analyze its high temperature limit. 
Section 4 is devoted to computing the holographic drag force using the string conjugate momenta in the Kerr-AdS black hole for both coordinate systems.
 In Section 5  the jet-quenching parameter is computed by means of a holographic light-like Wilson loop in  Kerr-${\rm AdS}_{5}$ with one rotation parameter in the static-at-infinity frame. Finally, Section 6 presents the conclusions. Some useful intermediate calculations are given in the Appendix. As indicated in some footnotes, this article is accompanied with SageMath notebooks.

\setcounter{equation}{0}
\section{The Setup}

\subsection{Open string dynamics}

Within the  holographic approach a heavy quark is described by  an open string moving in the black hole, which one endpoint is fixed on the spacetimes boundary. 
Thus, the momentum flowing from the boundary to the horizon is related to drag force applied on a heavy quark moving in the medium.
The equations of motion for the string can be obtained from the Nambu-Goto action that reads as
\be\label{SNGg}
S_{NG} = - \frac{1}{2\pi\alpha'} \int  d \sigma^{0} d\sigma^{1}\sqrt{-g},
\ee
where $\sigma^{0}, \sigma^{1}$ are coordinates on the string worldsheet which parametrize the induced metric, $g =\det g_{\alpha\beta}$ is the determinant of the induced metric, given by
\be\label{indm-h}
g_{\alpha\beta} = G_{MN}\partial_{\alpha}X^{M}\partial_{\beta}X^{N},
\ee
with the 5d background metric  $G_{MN}$ and the embedding functions of the worldsheet with $X^{\mu} = X^{\mu}(\sigma)$, $\sqrt{\alpha'}$ is a fundamental string scale.

Varying the action (\ref{SNGg}) with respect to $X^{\mu}$ we get the  equations of motion 
\be\label{maineqC1}
\frac{1}{\sqrt{-g}} \partial_\alpha \Big(\sqrt{-g}\, G_{\mu\nu} \,
 \partial^{\alpha} X^\nu\Big)-\frac12 \partial _\mu G_{\rho\nu}\partial_\alpha X^\rho\, \partial^{\alpha}X^\nu=0.\ee
The conserved currents are given through the variational derivatives on $\partial_{\alpha} X^{\mu}$
\be\label{conjmomenta}
\pi^{\alpha}_{\mu} \equiv -2\pi \alpha' \frac{\delta S}{\delta \partial_{\alpha} X^{\mu}} = \sqrt{-g} g^{\alpha \beta} G_{\mu\nu}(X)\partial_{\beta}X^{\nu}.
\ee
The energy density of the string is
\be\label{conenden}
\pi^{\sigma^{0}}_{t} =\sqrt{-g}\,g^{0\beta}G_{t\nu}(X) \,\partial_\beta X^\nu,
\ee
in what follows,  we keep for the conjugate momenta $\pi^{\sigma^{0}}=\pi^{0}$,  $\pi^{\sigma^{1}}=\pi^{r}$.
Therefore the total energy of the string is given by
\be\label{el}
E=  - \frac{1}{2\pi \alpha'}\int dr \pi^{0}_{t}.
\ee
Knowing the energy of the string one can calculate the mass of a single heavy quark and corrections to it. 
The energy (\ref{el}) should be interpreted as the free energy of a static quark sitting in the $\mathcal{N} =4$ SYM quark-gluon-plasma.
Another quantity of our interest is the drag force acting on a heavy quark.
The time-independent force on the string is
\be\label{dragforce}
\frac{\partial p_{\mu}}{\partial \sigma^{0}} = - \frac{1}{2\pi \alpha'} \pi^{r}_{\mu}.
\ee
The  force that acts on the string is calculated as follows
\be\label{dragfp}
\frac{\partial p_{\mu}}{\partial \sigma^{0}} = - \frac{1}{2\pi\alpha'}\pi^{r}_{\mu}.
\ee

\subsection{The 5d Kerr-AdS black hole in different coordinates}

Here we briefly describe the 5d Kerr-AdS black hole with two non-zero rotational parameters.
As it is known a rotating black hole in five dimensions can be characterized by the mass and two angular momentum invariants related to the numbers of Casimir invariants of $SO(4)$, therefore the solution has two rotational parameters.

The metric of the 5d Kerr-AdS solution \cite{HHT} in terms of the Boyer-Lindquist coordinates is given by
\bea\label{2}
ds^{2}& =& - \frac{\Delta_r}{\rho^{2}}\left(dt - \frac{a\sin^{2}\theta}{\Xi_{a}}d\phi - \frac{b\cos^{2}\theta}{\Xi_{b}}d \psi \right)^{2} + \frac{\Delta_{\theta}\sin^{2}\theta}{\rho^{2}}\left(adt - \frac{(r^{2} +a^{2})}{\Xi_{a}}d\phi \right)^{2}\nonumber\\
&+& \frac{\Delta_{\theta}\cos^{2} \theta}{\rho^{2}}\left( bdt  - \frac{(r^{2} +b^{2})}{\Xi_{b}}d\psi \right)^{2} + \frac{\rho^{2}}{\Delta_r}dr^{2} + \frac{\rho^{2}}{\Delta_{\theta} }d\theta^{2} \\
&+&\frac{(1 + r^{2}\ell^{2})}{r^{2}\rho^{2}}\left(ab dt -  \frac{b(r^{2} + a^{2 })\sin^{2}\theta}{\Xi_{a}} d\phi  -\frac{a(r^{2} + b^{2 })\cos^{2}\theta}{\Xi_{b}}d\psi \right)^{2},\nonumber
\eea
where $M$ is the mass of the black hole, $a$, $b$ are rotating parameters, constrained such that $a^{2}$, $b^{2}\leq \ell^{-2}$ and we also denote
\bea 
\Delta_r &=& \frac{1}{r^{2}}(r^{2} + a^{2})(r^{2} + b^{2})(1+ r^{2}\ell^{2}) -2M,\nonumber\\ \label{Xi.ab}
\Delta_{\theta} &=& (1- a^{2}\ell^{2}\cos^{2} \theta -b^{2} \ell^{2}\sin^{2}\theta), \\ 
\rho^{2}& =& (r^{2} + a^{2}\cos^{2}\theta + b^{2}\sin^{2}\theta),\nonumber \\ 
\Xi_{a}& = &(1 - a^{2} \ell^{2}), \quad \Xi_{b} = (1-b^{2}\ell^{2}).\nonumber
\eea
We note that for the spherical part of the metric we use Hopf coordinates, thus the angular coordinates are defined as  $0\leq\phi,\psi\leq 2\pi$, $0\leq\theta\leq\pi/2$. 
The horizon position is defined as a largest root $r_{+}$ to the equation $\Delta_{r} = 0$. So the solution (\ref{2}) is defined on the region of the radial coordinate $r\in (r_{+}, +\infty)$, with the boundary of Kerr-$AdS_{5}$ at $r = +\infty$. Note, that  the Boyer-Lindquist coordinates correspond to the rotating-at-infinity frame.

The conformal metric on the boundary with $r\to +\infty$ is represented by
\be\label{rotb}
ds^{2}_{BL} = -dt^{2} + \frac{2a\sin^{2}\theta}{\Xi_{a}}dtd\phi + \frac{2b\cos^{2}\theta}{\Xi_{b}}dtd\psi + \frac{\ell^{2}}{\Delta_{\theta}}d\theta^{2} + \frac{\ell^{2}\sin^{2}\theta}{\Xi_{a}}d\phi^{2} +
\frac{\ell^{2}\cos^{2}\theta}{\Xi_{b}}d\psi^{2}.
\ee
The Hawking temperature reads
\be\label{THawk}
T_{H}= \frac{1}{2\pi}\left(r_{+}(1+ r^{2}_{+}\ell^{2})\Bigl(\frac{1}{r^{2}_{+} + a^{2}} + \frac{1}{r^{2}_{+} + b^{2}}\Bigr) - \frac{1}{r_{+}}\right).
\ee
It worth to be noted that it corresponds to the temperature of the dual field theory. We see that for a theory that lives on the boundary of the Kerr-AdS spacetime with the metric (\ref{rotb}) and a temperature (\ref{THawk}) the Ehrenfest–Tolman  relation holds\footnote{We thank V. Braguta for pointing this.}
\be
\sqrt{|G_{00}|}T_{H} =\frac{1}{\beta} = \textrm{const}.
\ee
Plots with the dependence of the Hawking temperature on the horizon $r_{+}$ were presented in \cite{AGG2020}, 
where it was also shown that a Hawking-Page phase transition in the 5d Kerr-AdS black hole appears for $a$ and $b$ in the range of  values from $0$ to $0.25$ with $\ell =1$.

We note that in the Boyer-Lindquist coordinates, the metric is asymptotic to $AdS_{5}$ in a rotating frame, with the angular velocities
\be\label{Omegaab}
\Omega^{\infty}_{a} = a\ell^{2},\quad\Omega^{\infty}_{b} = b\ell^{2}.
\ee
Surprisingly, that if one takes into account that the rotational parameters have the dimension of  length, $\ell$ has the dimension of inverse length, we come that  $\Omega_{a,b}^{\infty}$ are measured in $\textrm{fm}^{-1}$ and coming to MeV,  we get the following range for the angular velocities 
\be
\Omega^{\infty}_{a,b} \approx 19.7 - 45.3 \quad \textrm{MeV}, 
\ee
that is in agreement with values of the angular  velocity that is used in the hydrodynamical simulations  \cite{QGPR2}.

The 5d Kerr-AdS black hole can be also represented  in asymptotically AdS coordinates.
Here it is convenient to take $\ell = 1$, then the coordinate transformations are
\bea\label{coordAdS}
(1 - a^{2})y^{2}\sin^{2}\Theta& = &(r^{2} + a^{2})\sin^{2}\theta \nonumber ,\\ 
(1 - b^{2})y^{2}\cos^{2}\Theta &=& (r^{2} + b^{2})\cos^{2}\theta,\nonumber\\ \label{coordAdS2}
\Phi &= &\phi + a t, \\
\Psi&= & \psi  + b t,\nonumber\\
T &= &t.\nonumber
\eea
For generic values of the rotational parameters $a$ and $b$,we are not able to write down the metric in AdS coordinates. So, in what follows, we will use the approximate form from \cite{Gibbons:2004ai}.
 The line-element of the 5d Kerr-AdS black hole in the coordinates (\ref{coordAdS})-(\ref{coordAdS2}) near the spacetime boundary $y\to + \infty$  reads
	\begin{eqnarray}\label{metlim}
ds^{2}&\simeq& -(1+y^{2})dT^{2} + \frac{dy^{2}}{1+y^{2} - \frac{2M}{\Delta^{2}y^{2}}} + y^{2}(d\Theta^{2} + \sin^{2}\Theta d\Phi^{2} + \cos^{2}\Theta d\Psi^{2})\nonumber\\
&+&\frac{2M}{\Delta^{3}y^{2}}dT^{2} + \frac{2M a^{2}\sin^{4}\Theta}{\Delta^{3}y^{2}}d\Phi^{2} + \frac{2Mb^{2}\cos^{4}\Theta}{\Delta^{3}y^{2}}d\Psi^{2}-\\
&-&\frac{4Ma\sin^{2}\Theta}{\Delta^{3}y^{2}}dTd\Phi - \frac{4Mb\cos^{2}\Theta}{\Delta^{3}y^{2}}dTd\Psi + \frac{4Mab\sin^{2}\Theta\cos^{2}\Theta}{\Delta^{3}y^{2}}d \Phi d\Psi,\nonumber
	\end{eqnarray}
where 
\be\label{deltaAdSc}
\Delta = 1 - a^{2}\sin^{2}\Theta -b^{2}\cos^{2}\Theta.\ee
Note, that the metric (\ref{metlim}) with $M=0$ describes the 5d AdS spacetime in the global coordinates.

The boundary of (\ref{metlim}) doesn't rotate at infinity (static-at-infinity frame) and has the following form
\be\label{nonrotb}
ds^{2} = - dT^{2} + d\Theta^{2} +\sin^{2} \Theta d\Phi^{2} + \cos^{2}\Theta d \Psi^{2},
\ee
which describes a compactified Minkowski space and manifests $\mathbb{R}\times\mathbb{S}^{3}$ symmetry.
It worth to be noted that the boundary of the Kerr-$AdS_{5}$ in Boyer-Lindquist coordinates for small rotational parameters 
$a$ and $b$ is approximately (\ref{nonrotb}). In \cite{Gibbons:2004ai} it was shown that the metrics of the boundary in BL coordinates and in static-at-the-boundary coordinates are conformally related as
\be
ds^2_{BL,bnd}  = \frac{y^2}{r^2} ds^2_{AdS,bnd}.
\ee  
So we can assume the Kerr-AdS black hole in BL coordinates is holographically dual to CFT at finite temperature in a rotating frame.

\setcounter{equation}{0}
\section{Energy of a static heavy quark}

In this section we compute the thermal mass of a heavy quark and find how the rotation affects on it.
 
A static quark at rest in $\mathcal{N}=4$ SYM quark-gluon plasma can be described by a straight string stretched from the spacetime boundary to the
black hole horizon. The solutions to the string equations of motion that represent this configuration are just constants. 
For the string in the Kerr-$AdS_5$ background in rotating-at-infinity frame (\ref{2})-(\ref{Xi.ab}) and with the worldsheet parametrization  $(\sigma^{0}, \sigma^{1}) = (t,r)$, it is given by
\be\label{ststBL}
\theta(r,t)=\theta_{0},\quad \phi(t,r)= \phi_{0}, \quad \psi(t,r) =\psi_{0}.
\ee 

To calculate the energy of the static quark one needs to find the energy of the string (\ref{ststBL}) defined by (\ref{el}).
To find this we need to calculate the conjugate momentum representing the energy density of a string. In the Boyer-Lindquist coordinates it has the following form 
\be\label{pitt}
\pi^{0}_{\,\,\,t} = \sqrt{-g}\left(g^{00}(G_{tt}+G_{t\phi}\dot{\phi}+G_{t\psi}\dot{\psi} )+g^{01}\left(G_{t\phi}\phi'+G_{t\psi}\psi' \right)\right).
\ee 
Plugging (\ref{ststBL}) into (\ref{pitt}) we get
\be\label{estring}
\pi^{0}_{\,\,\,t}= \sqrt{-g}g^{00}G_{tt} = - \left(1- \frac{\Delta_{\theta}}{\Delta_r}\left(a^2\sin^2\theta+b^2\cos^2\theta\right)-\frac{1}{\Delta_r}\frac{a^2b^2(1+r^2\ell^2)}{r^2}\right).
\ee
Expanding (\ref{estring})  first in small $a$, then in small $b$ and neglecting cross-terms of higher order  we get:
\be\label{expanestr}
\pi^{0}_{\,\,\,t}=-1+\frac{a^2 \sin ^2\theta + b^2 \cos ^2\theta}{\ell^2 r^4+r^2-2 M}.
\ee
It's interesting to note that in (\ref{expanestr}) there is a contribution related to the rotational parameters. Another interesting thing, observed in (\ref{expanestr}), is that the expression has a dependence on the blackening factor $\ell^2 r^4-2 M+r^2$ of the Schwarzchild-AdS black hole. This happens  since  the Kerr-AdS solution with small rotating parameters tends to the Schwarzchild-AdS  one.
Substituting (\ref{expanestr}) into (\ref{el}) and integrating,  we have
\bea
E&=&\frac{1}{2\pi \alpha'}\Bigl(r+\frac{ \left(a^2\sin^2\theta+b^2\cos^2\theta\right)}{2\sqrt{M}\sqrt{8 M\ell +1}}\Bigl(\sqrt{1+\sqrt{8 M\ell +1}} \tanh^{-1}\Bigl(\frac{\sqrt{2}\ell r }{\sqrt{\sqrt{8 M\ell+1}-1}}\Bigr) \nonumber\\
&-&\sqrt{\sqrt{8 M\ell +1}-1} \tan^{-1}\Bigl(\frac{\sqrt{2}\ell r }{\sqrt{1+\sqrt{8 M\ell +1}}}\Bigr)\Bigr)\Bigr)\Big|^{r_{m}}_{r_{H}+\epsilon},\nonumber
\eea
where we define the quark location by $r_{m}$.

At zero temperature limit, that corresponds to $M = 0$, we get
\bea
E\Big|_{T=0}
&=&\frac{1}{2\pi\alpha'}\left(r+\sqrt{2}\ell\left(a^2 \sin ^2\theta +b^2 \cos ^2\theta \right)\left(\frac{1}{\sqrt{2} \ell r}-\frac{\tan ^{-1}(\ell r)}{\sqrt{2}}-\frac{i \pi
}{4 \ell \sqrt{M}}\right)\right)\Big|^{r_m}_{r_H+\epsilon}=\nonumber\\
&=&\frac{1}{2\pi\alpha'}\left(r_m+\ell(a^2\sin^2\theta+b^2\cos^2\theta)\left(\frac{1}{\ell r_m}-\frac{1}{\varepsilon}-\tan^{-1}(\ell r_m)\right)\right),
\eea
where we use $\varepsilon=\ell\epsilon$ and $r_H=0$ with $M=0$. 
Note, that at $T=0$ the renormalized energy equals to the (Lagrangian) mass $m$ of the quark at zero temperature limit. So we have
\be
E_{\rm{ren}}\Big|_{T=0} = \frac{1}{2\pi\alpha'}\left(r_m+\ell(a^2\sin^2\theta+b^2\cos^2\theta)\left(\frac{1}{\ell r_m}-\tan^{-1}(\ell r_m)\right)\right) = m.
\ee
If one increases the temperature the energy takes the form 
\be
E = \frac{1}{2\pi \alpha'}\left(m -r_{H}- \Bigl(\frac{ \tanh^{-1}\Bigl(\frac{r_{H}+\epsilon}{r_{H}}\Bigr)}{r_{H}} - \ell \frac{\tan^{-1}\Bigl(\frac{\ell (r_{H}+\epsilon)}{\sqrt{\ell^2 r^{2}_{H}+ 1}}\Bigr)}{\sqrt{\ell^2 r^{2}_{H} + 1}}\Bigr)\frac{(a^2  \sin^{2}\theta+b^2\cos^2\theta)}{(2\ell^2 r^2_{H} + 1)}\right).
\ee
Therefore, the static thermal mass is given by
\be
M_{\rm{rest}} =  m - \Delta m(T,a,b),
\ee
where
\bea\label{deltamrotf}
\Delta m(T,a,b)  &=&   \frac{\sqrt{\lambda}}{2\pi}\Bigl(\frac{\pi T_{H} + \sqrt{\pi^2 T^{2}_{H}- 2\ell^2}}{2\ell^2}+\frac{\ell^2 (a^2  \sin^{2}\theta +b^2\cos^2\theta)}{\pi^2 T^2_{H} +  \pi T_{H}\sqrt{\pi^2 T^2_{H} - 2\ell^2}}\\
&\times& \Bigl(\frac{ \tanh^{-1}\Bigl(1+\frac{\epsilon2\ell^2}{\pi T_{H}+\sqrt{\pi^2 T^2_{H} - 2\ell^2}}\Bigr)}{\pi T_{H} + \sqrt{\pi^2 T^{2}_{H}- 2\ell^2}}-\frac{ \tan^{-1}\Bigl(\frac{ (\pi T_{H} + \sqrt{\pi^2 T^2_{H} - 2\ell^2}+2\epsilon\ell^2)}{\sqrt{2}\sqrt{\pi^2 T^{2}_{H}+\ell^2+ \pi T_H\sqrt{\pi^2T^2_{H} -2\ell^2}}}\Bigr)}{\sqrt{2}\sqrt{\pi^2 T^2_H + \ell^2 + \pi T_H\sqrt{\pi^2 T^2_{H} - 2\ell^2}}}
\Bigr)\Bigr),\nonumber
\eea
with $\lambda = 1/\alpha'^2$.

Note, that in the limit of the large $T_{H}$ the contribution from rotation is suppressed and comes to
\be\label{DeltaHT}
\Delta m(T,a,b)  \approx \frac{\sqrt{\lambda}}{2}\frac{ T_H}{\ell^2}.
\ee
This is exactly that was found for the planar AdS black holes  in \cite{HKKKY}.

Now we turn to the calculations in the  AdS coordinates (non-rotating-at-infinity frame) (\ref{metlim}). Using a similar worldsheet parametrization
$(\sigma^{0},\sigma^{1}) = (T,y)$ and considering a quark as an endpoint of a static string, we have for the conjugate momenta
\be\label{Pi0adsc}
\pi^{0}_{\,\,\,t}= G_{yy}G_{TT}=-1+\frac{2M(a^{2}\sin^{2}\Theta+b^{2}\cos^2\Theta)}{\Delta(\Delta^{2}y^{2}+\Delta^{2}y^{4} -2M)},
\ee
where $\Delta$ is given by (\ref{deltaAdSc}).

Taking into account that rotational parameters $a$ and $b$ are small,  we get from (\ref{Pi0adsc}) 
\be
\pi^{0}_{\,\,\,t}= -1 + \frac{2 M(a^2 \sin^2\Theta + b^2 \cos^2 \Theta) }{ y^4 + y^2 -2 M}.
\ee
Correspondingly, for the energy (\ref{el}) we have:
{\small
\bea\label{stqEAdSc}
E&=&\frac{1}{2\pi \alpha'}\Bigl(y+\frac{\sqrt{ M} \left(a^2\sin^2\Theta+b^2\cos^2\Theta\right)}{\sqrt{8 M+1}}\times\\
&\times&\Bigl(\sqrt{1+\sqrt{8 M+1}} \tanh^{-1}\Bigl(\frac{\sqrt{2}y }{\sqrt{\sqrt{8 M+1}-1}}\Bigr)-\sqrt{\sqrt{8 M+1}-1} \tan^{-1}\Bigl(\frac{\sqrt{2}y }{\sqrt{1+\sqrt{8 M+1}}}\Bigr)\Bigr)\Bigr)\Big|^{y_{m}}_{y_{H}+\epsilon}.\nonumber
\eea}

\begin{figure}[t]
  \centering
\includegraphics[scale=0.5]{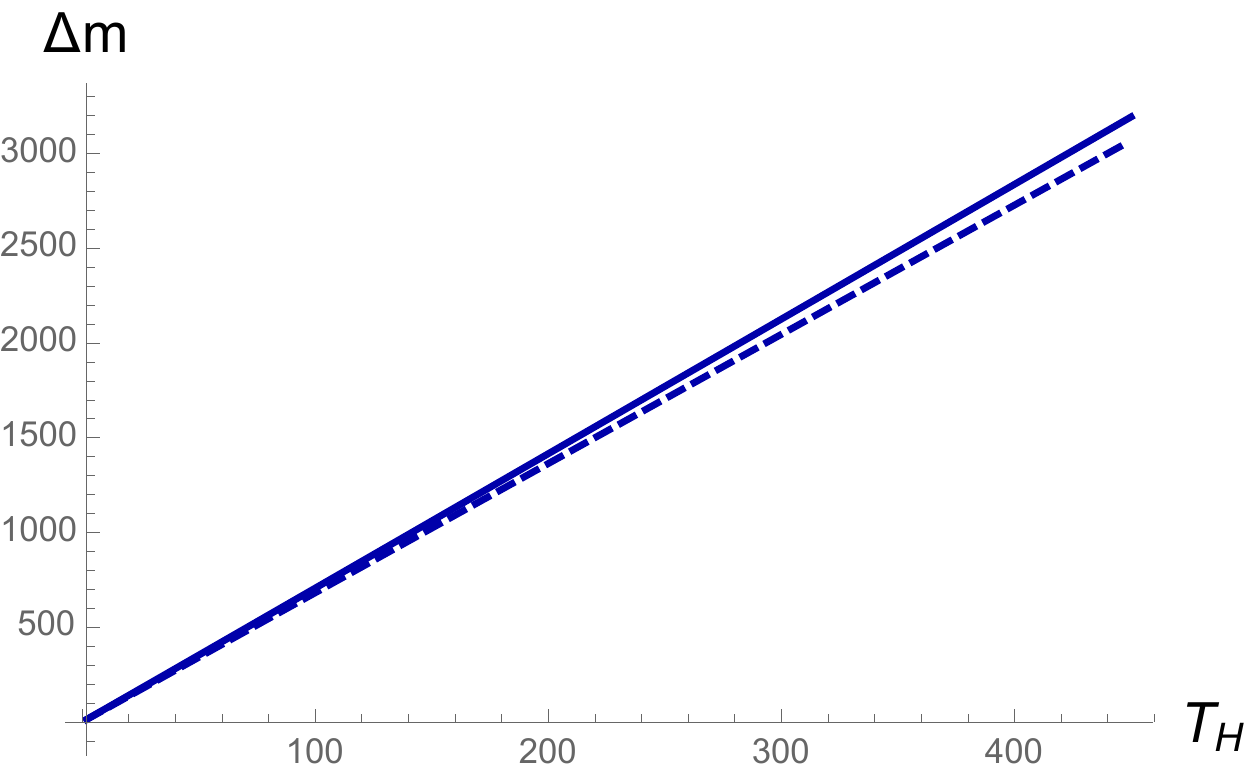}
 \caption{The dependence of $\Delta m$ (in the non-rotating-at-infinity-frame (\ref{deltamrotf}) and in the static-at-infinity-frame (\ref{mTabAdsC})) on the temperature $T_{H}$.}
  \label{Fig:deltam}
\end{figure}

Note, that in (\ref{stqEAdSc}) we again see terms related with rotation. However,
in the limit of zero mass, that corresponds to the zero temperature, the contribution from rotation vanishes:
\be\label{ET0AdS}
E|_{T=0}=\frac{1}{2\pi \alpha'} \left(y+2  \left(\frac{M}{ y}- M \tan ^{-1}y-\frac{i \pi  \sqrt{M}}{2\sqrt{2}} \right) \left(a^2 \sin ^2\Theta +b^2 \cos ^2\Theta \right)\right)\Big|^{y_m}_{y_{H}+\epsilon}=y_m,
\ee
where we also take into account that $y_{H}$ tends to zero as $M=0$.  This can be explained by that we consider the Kerr-AdS black hole in the non-rotating-at-infinity frame.

 As in the case previous case this energy (\ref{ET0AdS}) corresponds to the mass of the quark at rest:
\be\label{mAdSc}
E|_{T=0}=\frac{1}{2\pi\alpha'}y_m=m.
\ee
Correspondingly, the total energy with a switched temperature  taken into account is given by
\bea
E&=&\frac{1}{2\pi \alpha'}\Bigl(m-y_H - \Bigl(\sqrt{( y_{H}^2 +1)} \tanh ^{-1}\Bigl(1 +\frac{\epsilon }{y_H}\Bigr)-y_{H} \tan^{-1}\Bigl(\frac{y_H+\epsilon }{\sqrt{y^2_H+1}}\Bigr)\Bigr)\times \nonumber\\
&\times&\frac{y_{H}(\sqrt{y^2_{H} + 1})(a^2\sin^2\Theta+b^2\cos^2\Theta)}{2y^2_H+1}\Bigr),
	\eea
where we used the relation between $y_{H}$ and $M$, namely,  $M = \frac{y^2_{H}(y^2_{H} + 1)}{2}$.
Finally, we can represent the thermal mass of the quark at rest as
\be
M_{\rm{rest}} = m - \Delta m(T,a,b) ,
\ee
where $m$ is given by (\ref{mAdSc}) and
\bea\label{mTabAdsC}
\Delta m(T,a,b)  &=& 
  \frac{\sqrt{\lambda}}{2\pi}\Bigl(\frac{\pi T_{H} + \sqrt{\pi^{2}T^{2}_{H} -2}}{2} + \\
&+& \Bigl(\frac{2\tanh^{-1}\Bigl(1 + \frac{2\epsilon}{\pi T_{H} + \sqrt{\pi^2 T^2_{H} -2}}\Bigr) }{\pi T_{H}+\sqrt{\pi^2 T^2_{H}- 2}}
-\frac{\sqrt{2}\tan^{-1}\Bigl(\frac{1}{\sqrt{2}}\frac{\pi T_{H} + \sqrt{\pi^2 T^2_{H} - 2} +2\epsilon}{\sqrt{\pi^{2}T^{2}_{H} + \pi T_{H}\sqrt{\pi^2 T_{H}^2 -2} +1 }}\Bigr)}{\sqrt{\pi^2 T^{2}_{H} + \pi T_{H}\sqrt{\pi^2 T^{2}_{H}-2}+1}}\Bigr)\nonumber\\
&\times&\frac{\Bigl((\pi^2 T^2_{H} + \pi T_{H}\sqrt{\pi^2 T^2_{H} - 2})^2 - 1\Bigr)(a^2\sin^2\Theta + b^2 \cos^2\Theta)}{4(\pi^2 T^{2}_{H} + \pi T_{H}\sqrt{\pi^2 T^2_{H} -2})}\Bigr).\nonumber
\eea

From the relation (\ref{mTabAdsC})  we see, that at high temperatures the rotation doesn't affect on the thermal mass of the static quark, moreover the approximate form of the correction looks like for the  AdS black brane with a planar horizon \cite{HKKKY} and represents a linear dependence on $T_{H}$
\be
\Delta m(T,a,b)  \approx \frac{\sqrt{\lambda}}{2} T_H.
\ee
The same dependence for $\Delta m(T,a,b)$ was observed for the Kerr-$AdS_{5}$ in the rotating-at-infinity frame (\ref{DeltaHT}).
We compare the behaviors of $\Delta m(T,a,b)$ for different frames in fig.~\ref{Fig:deltam}.

\setcounter{equation}{0}
\section{Drag force acting on a heavy quark in 5d Kerr-AdS black hole}
\subsection{Calculations in Boyer-Lindquist coordinates}
In this section we consider a trailing curved string in the 5d Kerr-AdS black hole in Boyer-Lindquist coordinates (\ref{2}) with (\ref{Xi.ab}) (the rotating-at-infinity frame)\footnote{See the SageMath notebook \url{https://cocalc.com/share/6850015be0320058cadba1ce63e7a3eedf4eef89/Kerr-AdS-5D-string-b_na.ipynb} for some details of the computations.}.
We parametrize the string worldsheet as $\sigma^{0}=t$,  $\sigma^{1} = r$. For the embedding of the string worldsheet into the black hole background (\ref{2}) we choose
\be\label{coordpar}
\theta= \theta(t,r),\quad \phi = \phi(t,r), \quad \psi = \psi(t,r).
\ee

Owing to (\ref{coordpar}) the components of the induced metric (\ref{indm-h}) are 
\bea\label{h00}
g_{00} &= & G_{tt} + G_{\theta\theta}\dot{\theta}^2 + G_{\phi\phi}\dot{\phi}^2 + G_{\psi\psi}\dot{\psi}^2 + 2 \left(G_{t\phi}\dot{\phi} + G_{t\psi}\dot{\psi} + G_{\psi\phi}\dot{\psi}\dot{\phi}\right) ,\\
g_{01} &= &g_{10}= G_{\theta\theta}\dot{\theta}\theta' + G_{\phi\phi}\dot{\phi}\phi' + G_{\psi\psi}\dot{\psi}\psi' + G_{t\phi}\phi'+G_{t\psi}\psi'+G_{\phi\psi}\left(\dot{\phi}\psi'+\phi'\dot{\psi}\right),\\ \label{h11}
g_{11} &= & G_{\theta\theta}\theta'^2 + G_{\phi\phi}\phi'^2+ G_{\psi\psi}\psi'^2+G_{rr}+ 2 G_{\psi\phi}\psi'\phi',
\eea
where we denote $\dot \,\, = \frac{d}{dt}$ and $' = \frac{d}{dr}$.

 The equations of motion (\ref{maineqC1}) with (\ref{h00})-(\ref{h11}) are difficult to work with,  so we are going to consider the linearized equations, which we obtain using an expansion in small rotating parameters.
We are  looking for  solutions to eqs.(\ref{maineqC1}) in the form of the curved string solution given by the ansatz 
\bea
\phi(t,r) &=& \Phi_{0} + a\ell^{2}t  +  a\phi_{1}(r) + \mathcal{O}(a^{2}),\quad \nonumber\\\label{cs3}
\theta(r)  &= &\Theta_{0} +( a+b)^{2}\theta_{1}(r) + \mathcal{O}((a+b)^{4}),\\ 
\psi(t,r) &= &\Psi_{0} + b\ell^2 t +b\psi_1(r) + \mathcal{O}(b^{2}), \quad \nonumber
\eea
where $\Theta_{0},\Phi_{0},\Psi_{0}$ are some constants,  $\theta_{1},\phi_{1},\psi_{1}$ are some functions depending on the radial coordinate $r$.  We recall that the string endpoint attached to the boundary is fixed and rotates with the thermal medium. So the terms $a\ell^2$, $b\ell^2$ can be associated to the components of the angular velocity of the boundary.
To simplify calculations we also assume that the following relation between the angular parameters hold
\be\label{bna} 
b= n a,  
\ee 
where $n$ is some  number thus the parameters $a$ and $b$ are of the same order.

The constraint (\ref{bna}) allows to reduce the calculations in terms of one parameter of expansion $a$.  Plugging (\ref{cs3}) with (\ref{bna}) into (\ref{h00})-(\ref{h11}) 
 the determinant of the induced metric takes the following form
\bea\label{indm-coeff}
- g&=& \left( a^2\ell^2(G_{\phi\phi}\phi_1' + n^2 G_{\psi\psi}\psi_1' ) + a(G_{t\phi}\phi_1' + n  G_{t\psi}\psi_1')+ n a^2 \ell^2 G_{\phi\psi}\left(\psi_1'+\phi_1'\right)\right)^2 \nonumber\\
 &-& \left( G_{tt} + a^2\ell^4G_{\phi\phi} + n^2a^2 \ell^4G_{\psi\psi} + 2\ell^2 \left(a G_{t\phi} + n a G_{t\psi} + n a^2\ell^2 G_{\psi\phi}\right)\right) \\
 &\times&\left((1+n)^2a^2G_{\theta\theta}\theta_1'^2 + a^2 G_{\phi\phi}\phi_1'^2+ n^2 a^2 G_{\psi\psi}\psi_1'^2+G_{rr}+ 2n a^2 G_{\psi\phi}\psi_1'\phi_1'\right),\nonumber
\eea
with the metric components  (\ref{metriccomp})-(\ref{metriccomp2}).

Thanks to the ansatz (\ref{cs3}) the equations of motion for $\theta,\phi,\psi$ comes to the equations for $\theta_{1}, \phi_{1}, \psi_{1}$. One can see that  $\phi_{1}$ and $\psi_{1}$ are cyclic variables, the corresponding conservation laws for $\phi_{1}$ and $\psi_{1}$ are given by
\be\label{pqcons}
\phi_{1} = \mathfrak{p} \int^{r}_{r_{+}} \frac{d\bar{r} }{\bar{r}^{2}+ \ell^2 \bar{r}^{4} -2M},\quad \psi_{1} = \mathfrak{q} \int^{r}_{r_{+}} \frac{d\bar{r}}{\bar{r}^{2}+ \ell^2 \bar{r}^{4} -2M}, 
\ee
where we used the conserved quantities $\mathfrak{p}, \mathfrak{q}$. Note, that we again observe the contribution of the Schwarzschild-AdS blackening functions in (\ref{pqcons}).  This appears because of small rotating parameters.  Plugging (\ref{pqcons}) into  (\ref{indm-coeff}) and doing expansion of the Lagrangian with (\ref{indm-coeff}) by $a$ up to 4-th order we get 
the equation for $\theta_{1}$, which can be represented as a first order differential equation, namely
\be\label{diffeqtheta}
\Upsilon' +\frac{2r + 4\ell^2r^3}{r^{2} + \ell^2 r^{4} -2M}\Upsilon +\frac{1+4\ell^2M(n^2-1)-\mathfrak{p}^{2}+n^{2}(\mathfrak{q}^{2}-1)-3(n^{2} -1)\ell^4r^{4}}{2 (1+n)^{2}(r^{2} + \ell^2 r^{4}-  2M)^2}\sin(2\Theta_{0}) =0,
\ee
where we define $\Upsilon=\theta'_{1}$.

 The solution to (\ref{diffeqtheta}) is represented in the following form
\begin{eqnarray}\label{eqofM}
		\theta'_{1}& =&\frac{C_{1}}{r^{4}f(r)} -\frac{3\ell^{2}(1-n^{2})r\sin(2\Theta_{0})}{2(1+n)^{2}r^{4}f(r)}+\nonumber\\
		&+&\frac{(1-n^{2})\left(\ell^2r_H^2-1\right)^2 -\mathfrak{p}^{2} +n^{2}\mathfrak{q}^{2}}{2(n+1)^{2}f(r)r^{4}(2\ell^{2}r^{2}_{H} +1) r_{H}}\tanh^{-1}\left(\frac{r_{H}}{r}\right)\sin(2\Theta_{0})  \\
		&+&\ell\frac{(1-n^{2})\left(\ell^2r_H^2+2\right)^2 - \mathfrak{p}^2 + n^{2}\mathfrak{q}^{2}}{2(n+1)^{2}r^{4}f(r)(2\ell^{2}r^{2}_{H} +1)\sqrt{\ell^2 r^{2}_{H}+1}}\tan^{-1}\left(\frac{\ell r}{\sqrt{\ell^{2}r^{2}_{H} +1}}\right)\sin(2\Theta_{0}),\nonumber
	\end{eqnarray}
where $C_{1}$ is a constant of integration, $r_{H}$ is a horizon of the AdS-Schwarzchild black hole defined by (\ref{SchwAdShor}) and the function $f(r)$ is the blackening function of the  AdS-Schwarzchild black hole with a  spherical horizon given by
\be\label{fAdSch}
f(r) = \ell^2 + \frac{1}{r^2} - \frac{2M}{r^4}.
\ee 
We note that  $r_{H}$ is introduced in (\ref{eqofM}) recognizing some relations for the black hole mass (\ref{usrel4Mlm2})-(\ref{usrel4Ml3}).  The appearance of $f(r)$ (\ref{fAdSch}) and $r_{H}$ in (\ref{diffeqtheta})-(\ref{eqofM}) can be explained the rotational parameters are small and the leading order is related to the non-rotating case. 
Taking into account $n=b/a$ we represent (\ref{eqofM}) as follows
\begin{eqnarray}\label{eqofMab}
		\theta'_{1}& =&\frac{C_{1}}{r^{4}f(r)} -\frac{3\ell^{2}(a-b)r\sin(2\Theta_{0})}{2(a+b)r^{4}f(r)}+\nonumber\\
		&+&\frac{(a^2-b^{2})\left(\ell^2r_H^2-1\right)^2 -a^2\mathfrak{p}^{2} +b^{2}\mathfrak{q}^{2}}{2(a+b)^{2}f(r)r^{4}(2\ell^{2}r^{2}_{H} +1) r_{H}}\tanh^{-1}\left(\frac{r_{H}}{r}\right)\sin(2\Theta_{0})  \\
		&+&\ell\frac{(a^2-b^{2})\left(\ell^2r_H^2+2\right)^2 - a^2\mathfrak{p}^2 + b^{2}\mathfrak{q}^{2}}{2(b+a)^{2}r^{4}f(r)(2\ell^{2}r^{2}_{H} +1)\sqrt{\ell^2 r^{2}_{H}+1}}\tan^{-1}\left(\frac{\ell r}{\sqrt{\ell^{2}r^{2}_{H} +1}}\right)\sin(2\Theta_{0}).\nonumber
	\end{eqnarray}

The corresponding conjugate momenta (\ref{conjmomenta}) are calculated as
\bea\label{conjmom2}
\pi^{r}_{\theta} &=& f(r)r^{4}\theta'_{1}(a+b)^2 + \mathcal{O}((a+b)^{4}), \nonumber \\
\pi^{r}_{\phi} &= & f(r)r^{4}\phi'_{1}\sin^{2}(\Theta_{0})a + \mathcal{O}(a^{2}),\\
\pi^{r}_{\psi}& =&   f(r)r^{4}\psi'_{1}\cos^{2}(\Theta_{0})b + \mathcal{O}(b^{2}).\nonumber
\eea
Substituting (\ref{pqcons}) and (\ref{eqofM}) into  (\ref{conjmom2})  and expanding near the boundary $r\to +\infty$,  we obtain for the conjugate momenta
		\begin{eqnarray} 
	\pi^{r}_{\theta}&=&\Bigl(\frac{2C_1}{\sin (2 \Theta_{0} )}-3r\ell^2(1-n^{2})-\frac{3(1-n^2)}{r} \\ 
	&+&\frac{\ell\pi}{2(2\ell^{2}r^{2}_{H} +1)}\frac{\left(1-n^2\right)\left(\ell^2r_H^2+2\right)^2- \mathfrak{p}^2 +n^{2}\mathfrak{q}^{2}}{\sqrt{\ell^2 r^{2}_{H}+1}}\Bigr)\sin (2 \Theta_{0} )\frac{a^2}{2}+\mathcal{O}((a+b)^4),\nonumber\\
\pi^{r}_{\phi} &= & \mathfrak{p}\sin(\Theta_{0})^{2}a + \mathcal{O}(a^2),\quad \pi^{r}_{\psi} =  \mathfrak{q}\cos(\Theta_{0})^{2}b + \mathcal{O}(b^2).
	\end{eqnarray} 
	
The components of  the drag force  for $a\neq b$ with $b=na$  are
\begin{eqnarray}\label{dpdtab}
	\frac{d p_{\theta}}{dt}&=&\Bigl(-\frac{2C_1}{\sin (2 \Theta_{0} )}+3r\ell^2(1-n^{2})+\frac{3(1-n^2)}{r} \\ 
	&-&\frac{\pi\ell}{2(2\ell^{2}r^{2}_{H} +1)}\frac{\left(1-n^2\right)\left(\ell^2r_H^2+2\right)^2- \mathfrak{p}^2 +n^{2}\mathfrak{q}^{2}}{\sqrt{\ell^2 r^{2}_{H}+1}}\Bigr)\frac{\sin (2 \Theta_{0})}{4\pi \alpha'}a^2+\mathcal{O}((a+b)^4),\nonumber\\
\frac{d p_{\phi}}{dt}&=&-\frac{1}{2\pi\alpha'}\mathfrak{p}\sin(\Theta_{0})^{2}a + \mathcal{O}(a^2), \quad
 \frac{d p_{\psi}}{dt}=-\frac{1}{2\pi\alpha'}\mathfrak{q}\cos(\Theta_{0})^{2}b + \mathcal{O}(b^2)
\end{eqnarray}
or by virtue of (\ref{eqofMab}) we get for the $\theta$-component
\begin{eqnarray}
	\frac{d p_{\theta}}{dt}&=&\Bigl(-\frac{2C_1}{\sin (2 \Theta_{0} )}+3r\ell^2\frac{a-b}{a+b}+\frac{3}{r}\frac{a-b}{a+b} \\ 
	&-&\frac{\pi\ell}{2(2\ell^{2}r^{2}_{H} +1)}\frac{\left(a^2-b^2\right)\left(\ell^2r_H^2+2\right)^2- a^2\mathfrak{p}^2 +b^{2}\mathfrak{q}^{2}}{(a +b)^2\sqrt{\ell^2 r^{2}_{H}+1}}\Bigr)\frac{\sin (2 \Theta_{0})}{4\pi \alpha'}(a+b)^2+\mathcal{O}((a+b)^4).\nonumber
\end{eqnarray}

The linear divergent term (\ref{dpdtab}) is associated with the mass of the heavy quark, which can be renormalized by a cut-off $r_{m}$.  Taking  $r_{m} = 2\pi\alpha'M_{\rm rest}$ we get
\be\label{dpdtTH}
\displaystyle{\frac{dp_{\theta}}{dt}}= 3\frac{a-b}{a+b}\left(\ell^2 M_{\rm rest} + \frac{1}{4\pi\alpha'}\left(\pi T_{H} + \sqrt{\pi^2 T^{2}_{H}- 2\ell^2}\right) \right)\frac{(a+b)^{2}}{2}\sin(2\Theta_{0})
+...\quad.
\ee
Taking into account  (\ref{Omegaab})   we can also represent the relation for the drag force (\ref{dpdtTH}) with the contributions from the rotational velocities 
\be
\displaystyle{\frac{dp_{\theta}}{dt}}= 3\frac{(\Omega^{\infty}_a)^2-(\Omega^{\infty}_b)^2}{2\ell^2}\left( M_{\rm rest} + \frac{\sqrt{\lambda}}{4\pi \ell^2}\left(\pi T_{H} + \sqrt{\pi^2 T^{2}_{H}- 2\ell^2}\right) \right)\sin(2\Theta_{0})
+...\quad.
\ee
Consider $b= 0$, this corresponds to the case with the one non-zero rotational parameter \cite{AGG2020}, we have for the components of the drag force:
\begin{eqnarray}\label{dpthdt3}
	\frac{d p_{\theta}}{dt}&=&3\left(\ell^2 M_{\rm rest} + \frac{1}{4\pi\alpha'}\left(\pi T_{H} + \sqrt{\pi^2 T^{2}_{H}- 2\ell^2}\right) \right)\frac{a^{2}}{2}\sin(2\Theta_{0})
+...\quad,\\ \label{dpppdt3}
\frac{d p_{\phi}}{dt}&=&-\frac{1}{2\pi\alpha'}\mathfrak{p}\sin(\Theta_{0})^{2}a + \mathcal{O}(a^2), \quad \frac{d p_{\psi}}{dt}=0.
\end{eqnarray}
The relations (\ref{dpthdt3})-(\ref{dpppdt3}) are also in agreement with the predictions from the 4d Kerr-AdS black hole \cite{NAS}.  However, it worth to be noted that for two  arbitrary non-zero rotational parameters the coefficient for the leading term is defined by the combination of $a$ and $b$.

Another interesting particular case is the case of two equal rotational parameters, that corresponds to $n=1$. From (\ref{dpdtTH}) we see that the component of  the drag force $\frac{dp_{\theta}}{dt}$  just vanishes. This confirms  the result obtained in \cite{AGG2020} by using the metric with two equal  parameters.

\subsection{Calculations in asymptotically AdS coordinates}
Now we turn to the calculations of the drag force in the 5d Kerr-AdS black hole in asymptotically AdS coordinates (\ref{metlim}), i.e.  non-rotating-at infinity frame\footnote{See the SageMath notebook \url{https://cocalc.com/share/9e8e6273e8459a7ee654e124ec5ce50825096b2b/Kerr-AdS-5D-string-b_a-AdS.ipynb} for some details of the computations.}.
For the coordinates on the string worldsheet we take the parametrization $\sigma^{0}=T$, $\sigma^{1} =y$, which is similar to the previous case.

The embedding is defined through
\be\label{emdedads}
\Theta = \Theta(T,y), \quad \Phi = \Phi(T, y), \quad \Psi =\Psi(T,y).
\ee
By virtue of (\ref{emdedads}), the induced metric (\ref{indm-h}) have the following components:
\bea
g_{00}&=& G_{TT}+G_{\Theta\Theta}\dot{\Theta}^2+G_{\Phi\Phi}\dot{\Phi}^2+G_{\Psi\Psi}\dot{\Psi}^2+2\left(G_{T\Phi}\dot{\Phi}+G_{T\Psi}\dot{\Psi}+G_{\Phi\Psi}\dot{\Phi}\dot{\Psi}\right),\\ 
g_{01} &=& G_{\Theta\Theta}\dot{\Theta}\Theta'+G_{\Phi\Phi}\dot{\Phi}\Phi'+G_{\Psi\Psi}\dot{\Psi}\Psi'+G_{T\Phi}\Phi'+G_{T\Psi}\Psi'+G_{\Phi\Psi}(\Phi'\dot{\Psi}+\dot{\Phi}\Psi'),\\
g_{11}&=&G_{yy}+G_{\Theta\Theta}\Theta^{'2}+G_{\Phi\Phi}\Phi^{'2}+G_{\Psi\Psi}\Psi^{'2}+2G_{\Phi\Psi}\Phi'\Psi',
\eea
where we define  $\dot \,\, = \frac{d}{dt}$ and $' = \frac{d}{dr}$ and the components of the metric (\ref{metlim}) are given by (\ref{compAdS}).

As in the Boyer-Lindquist coordinates the equations of motion (\ref{maineqC1}) are complicated and we are going to look for the linearized solutions. Assuming that the rotational parameters are small and remembering the choice $\ell =1$ in this coordinates, the the curved string is given by the ansatz 
\bea\label{expg}
\Phi(T,y) &=& \Phi_{0} + aT  +  a\Phi_{1}(y) + \mathcal{O}(a^{2}),\quad \nonumber\\
\Theta(y)  &= &\Theta_{0} +\Theta_{1}(y)( a+b)^{2} + \mathcal{O}((a+b)^{4}),\\ \label{expg3}
\Psi(T,y) &= &\Psi_{0} + bT +b\Psi_1(y) + \mathcal{O}(b^{2}), \quad \nonumber
\eea
where $\Phi_{0}$, $\Theta_{0}$, $\Psi_{0}$ are some constants. Comparing to the previous case, the boundary of the background doesn't rotate, however we suppose that  the endpoint of the string slowly moves on the boundary.

With  (\ref{expg}) the determinant of the induced metric comes to
\bea\label{indmaads}
- g &=&\Bigl((a^{2}\sin^{2}\Theta \Phi'_{1} + b^{2}\cos^{2}\Theta \Psi'_{1})\Bigl(y^{2}+\frac{2M}{\Delta^{3}y^{2}}(a^{2}\sin^{2}\Theta+ b^{2}\cos^{2}\Theta -1 )\Bigr)\Bigr)^2\nonumber\\
& - &\Bigl((a^{2}\sin^{2}\Theta + b^{2}\cos^{2}\Theta-1)\Bigl(y^{2}+\frac{2M}{\Delta^{3}y^{2}}(a^{2}\sin^{2}\Theta + b^{2}\cos^{2}\Theta-1)\Bigr)-1\Bigr)\\
&\times&\Bigl(\frac{1}{1+y^{2} - \frac{2M}{\Delta^{2}y^{2}}} + y^{2}((a+b)^{4}\Theta'^{2}_{1}+a^{2}\sin^{2}\Theta\Phi'^{2}_{1} + b^{2}\cos^{2}\Theta\Psi'^{2}_{1})+\nonumber\\ 
&+ &\frac{2M}{\Delta^{3}y^{2}}(a^{2}\sin^{2}\Theta\Phi'_{1} +b^{2}\cos^{2}\Theta\Psi'_{1})^{2}\Bigr), \nonumber
\eea
where $\Theta$ is given by (\ref{expg}).  The form of the metric in the AdS coordinates (\ref{metlim}) is simpler than in the Boyer-Lindquist coordinates (\ref{2}), this leads to the relatively simple form of the induced metric (\ref{indmaads}). So, one doesn't need to put a constraint (\ref{bna}) we can do the expansion of the string Lagrangian first with respect to the parameter $a$ and  then with respect to $b$ removing the cross-terms.

From (\ref{indmaads}) we see that the equations of motion are reduced to EOM for $\Theta_{1}$,  $\Phi_{1}$ and $\Psi_{1}$. Moreover, 
for $\Phi_{1}$ and $\Psi_{1}$ we have the conservation laws  given by the following expressions:
\bea\label{conslawsads}
\Phi_{1} = \mathfrak{p}\int^{y}_{y_{+}}  \frac{d\bar{y}}{(\bar{y}^{2}+\bar{y}^{4} -2M)},\quad \Psi_{1} = \mathfrak{q}\int^{y}_{y_{+}} \frac{d\bar{y}}{(\bar{y}^{2}+\bar{y}^{4} -2M)},
\eea
where $\mathfrak{p}$, $\mathfrak{q}$ are constants.

Plugging (\ref{conslawsads}) into (\ref{indmaads}) and expanding the expression up to the 4th-order by $a$ and then by $b$, removing the cross-terms of higher orders, we get the equation of motion for $\Theta'_{1}$ 
\bea\label{thetaads}
\Upsilon'+\Upsilon\frac{2y\left(1+2y^2\right)}{-2M+y^2+y^4}+\frac{ (y^4-4M)(a^2 - b^2) -a^2\mathfrak{p}^2+b^2\mathfrak{q}^2}{2(a+b)^2\left(-2M+y^2+y^4\right)^2}\sin 2\Theta_0=0,
\eea
where $\Upsilon = \Theta'_{1}$.
Eq. (\ref{thetaads}) has the following solution:
\begin{eqnarray}
\Theta'_{1} &=&\frac{C_1}{y^4+ y^2-2 M}-\frac{y(a^2-b^2)}{2(a+b)^2(y^2+y^4-2 M)}\sin(2\Theta_{0})\\
&+&\frac{-(a^2-b^2)y_H^2(y^2_{H}+2) - a^2 \mathfrak{p}^2 + b^2 \mathfrak{q}^2 }{2y_H \left(2 y_H^2+1\right) (a+b)^2 (y^2+y^4-2 M)}\tanh ^{-1}\Bigl(\frac{y_{H}}{y}\Bigr) \sin(2\Theta_{0})+ \nonumber\\
&+&\frac{(a^2 -b^2)(1-y_H^4)-a^2\mathfrak{p}^2+b^2 \mathfrak{q}^2}{2 \sqrt{y_H^2+1} \left(2 y_H^2+1\right) (a+b)^2 (y^2+y^4-2 M)}\tan ^{-1}\Bigl(\frac{y}{\sqrt{y_H^2+1}}\Bigr) \sin(2\Theta_{0}), \nonumber
\end{eqnarray}
where $C_{1}$ is a constant of integration.

 The conjugate momenta (\ref{conjmomenta}) can be found as follows
\bea
\pi^{y}_{\Theta} &=&   (y^2+y^4-2M)\Theta'_{1}(a+b)^2+ \mathcal{O}((a+b)^{4}),\nonumber \\
\pi^{y}_{\Phi} &= & (y^2+y^4-2M)\sin^{2}(\Theta_{0})\Phi'_{1}a + \mathcal{O}(a^{2}),\\
\pi^{y}_{\Psi}& =&  (y^2+y^4-2M)\cos^{2}(\Theta_{0})\Psi'_{1}b + \mathcal{O}(b^{2}). \nonumber
\eea

The drag force components  (\ref{dragforce}) near the  boundary of Kerr-$AdS_{5}$, $y\to +\infty$,  are given by
\bea\label{dpdTheta}
\frac{dp_\Theta}{dt}&=& \Bigl(- \frac{2\tilde{C}_1}{\sin(2\Theta_{0})}+ \frac{y(a-b)}{(a+b)}+ \frac{(a-b)}{(a+b) y} \\ 
&-&\frac{\pi}{2}\frac{(a^2-b^2) (1-y_H^4)-a^2\mathfrak{p}^2 + b^2 \mathfrak{q}^2}{ \sqrt{y_H^2+1} (2 y_H^2+1)(a+b)^2 } \Bigr)\frac{(a+b)^2}{4\pi\alpha'}\sin(2\Theta_{0})+ \mathcal{O}((a+b)^{4}),\nonumber\\
\frac{dp_\Phi}{dt}&=&-\frac{\mathfrak{p}}{2\pi\alpha'}\sin^{2}(\Theta_{0})a + \mathcal{O}(a^{2}),\quad
\frac{dp_\Psi}{dt}=-\frac{\mathfrak{q}}{2\pi\alpha'}\cos^{2}(\Theta_{0})b  + \mathcal{O}(b^{2}).
\eea
As in the rotating-at-infinity frame we observe the divergent term with $y\to +\infty$ in (\ref{dpdTheta}), which can be associated to an infinite mass of the quark. Introducing a cut-off as  $y_{m} = 2\pi\alpha'M_{\rm rest}$ we get
\bea\label{dpdtAdSc}
\frac{dp_{\Theta}}{dt}&\approx& \frac{a-b}{a+b}\left( M_{\rm rest} + \frac{1}{4\pi\alpha'}\left(\pi T_{H} + \sqrt{\pi^2 T^{2}_{H}- 2}\right) \right)\frac{(a+b)^{2}}{2}\sin(2\Theta_{0})
+...,\\ \label{dpdtAdSc2}
\frac{d p_{\Phi}}{dt}&=&-\frac{1}{2\pi\alpha'}\mathfrak{p}\sin(\Theta_{0})^{2}a + \mathcal{O}(a^2), \quad \frac{d p_{\Psi}}{dt}=0.
\eea
Comparing (\ref{dpdtAdSc}) to (\ref{dpdtTH}) we see that  the form of the relations for the drag force in the rotating-at-infinity and static frames is similar up to a numerical coefficient that appears probably because of using an approximate form of the metric in static-at-infinity frame.

Consider interesting particular cases,  particularly, when we have just one non-zero rotational parameter,   we again obtain  a close relation but with another number coefficient,  but the dependence on the rotational parameters is exactly the same as in (\ref{dpthdt3})-(\ref{dpppdt3}).  As for two equal rotational parameters $a=b$ we obtain  that the drag force (\ref{dpdtAdSc}) is just zero, that matches with the results of the previous section and \cite{AGG2020}.

\setcounter{equation}{0}

\section{Jet quenching parameter}
Now we turn to computation of the jet-quenching parameter following up the holographic prescription from \cite{HoloJet}.
The jet-quenching parameter $\hat{q}$  is related with the  average of the light-like Wilson loop in the following form 
\be\label{JQPW}
W[C] = e^{-\frac{1}{4\sqrt{2}}\hat{q}L^{-}L^2},
\ee
where $C$ is a  contour,  $L^{-}$ is an extension in a light-like direction and   $L$ is an extension in a transversal one. In the holographic approach $W[C]$
is  equal to the  classical  string  action  stretched on the contour $C$ on the boundary of the holographic background
\be\label{JQPS}
W[C] = e^{-2S}.
\ee

We consider the Kerr-AdS  black hole in static-at-the-boundary frame (\ref{metlim}) with one non-zero rotational parameter $a\neq0$.  The metric reads 
	\begin{eqnarray}\label{metlimao}
ds^{2}&\simeq& -(1+y^{2})dT^{2} + \frac{dy^{2}}{1+y^{2} - \frac{2M}{\Delta^{2}y^{2}}} + y^{2}(d\Theta^{2} + \sin^{2}\Theta d\Phi^{2} + \cos^{2}\Theta d\Psi^{2})\\
&+&\frac{2M}{\Delta^{3}y^{2}}\Bigl(dT^{2} + a^{2}\sin^{4}\Theta d\Phi^{2} -2 a\sin^{2}\Theta dT d\Phi \Bigr),\nonumber
	\end{eqnarray}
where $\Delta = 1 - a^{2}\sin^{2}\Theta$.

The light-cone coordinates are defined by
\bea\label{lightcone}
dx^{+} &=& dT - a d\Phi, \quad  dx^{-} = dT+ a d\Phi,\\
dTd\Phi &=& \frac{1}{4a}\left((dx^{-})^2  -(dx^{+})^2\right), \quad dT = \frac{dx^{+} +dx^{-}}{2}, \quad d\Phi = \frac{dx^{-} -dx^{+}}{2a}. 
\eea
By owing of the coordinate transformation (\ref{lightcone}) the metric (\ref{metlimao}) takes the form\footnote{See the SageMath notebook \url{https://cocalc.com/share/f8db4d9b694aa3a8b5ac7b72f83aa370a257c9f1/Kerr-AdS-5D-asymptotic_b_0.ipynb}}:
\bea\label{lcKerrAdS5}
ds^{2}&\simeq& -\frac{1}{4}(dx^{+})^{2}\Bigl(1+y^{2}-\frac{y^{2}\sin^{2}\Theta}{a^{2}} -\frac{2M}{\Delta^{3}y^{2}}(1+\sin^{2}\Theta)^{2}\Bigr)\\
&-&\frac{1}{4}(dx^{-})^{2}\Big(1+y^{2}-\frac{y^{2}\sin^{2}\Theta}{a^{2}}-\frac{2M}{\Delta^{3}y^{2}}\cos^{4}\Theta\Bigr)\nonumber\\
&-&\frac{1}{2}dx^{+}dx^{-}\left(1+y^{2}+\frac{y^{2}\sin^2\Theta}{a^2} - \frac{2M}{\Delta^{3}y^{2}}(1-\sin^{4}\Theta )\right) \nonumber\\
&+& \frac{dy^{2}}{1+y^{2} - \frac{2M}{\Delta^{2}y^{2}}} + y^{2}(d\Theta^{2}  + \cos^{2}\Theta d\Psi^{2}).
\nonumber
\eea
The Wilson loop can be considered as an open string which both endpoints are attached to the boundary of the Kerr-$AdS_{5}$ background.  The string worldsheet coordinates are given by 
\be\label{paramone}
\sigma^{0}  = x^{-}, \quad \sigma^{1} = \Psi.
\ee
Along $\Psi$ we have length $L$ and the lightcone direction has a length $L^{-}$.
We suppose that 
\be\label{paramtwo}
x^{\mu} =x^{\mu}(\sigma^{1}).
\ee
The Wilson loop lies at constant $x^{+}$ and $\Theta$, then we have
\be
\Theta(\sigma^{1}) = \textrm{const}, \quad x^{+}(\sigma^{1}) = \textrm{const}.
\ee
For the holographic coordinate $y$ we impose the following constraint 
\be
y\Bigl(\pm  L/2\Bigr) = \infty,
\ee
thus we have
\be
y(\sigma^{1}) = y(-\sigma^{1}).
\ee
The Nambu-Goto action is given by (\ref{SNGg}) with the induced metric (\ref{indm-h}). By owing the parametrization (\ref{paramone})-(\ref{paramtwo})
the induced metric components are
\be
g_{00}=G_{x^{-}x^{-}},\quad g_{01}=0,\quad g_{11}=G_{yy}y'^2+G_{\Psi\Psi},
\ee
where we define by $A' =\frac{\partial A}{\partial \Psi}$.

According to these calculations the Nambu-Goto action has the following view:
{\small
\be\label{actionLC}
S=\frac{L^{-}}{2\pi\alpha'}\int^{L/2}_{-L/2}d\Psi\sqrt{\frac{1}{4}\Bigl(y^{2}\cos^{2}\Theta + \frac{y'^{2}}{1+y^{2}- \frac{2M}{y^{2}\Delta^2}}\Bigr)\left(1+y^{2}-\frac{y^{2}\sin^{2}\Theta}{a^2} -\frac{2M\cos^4\Theta}{\Delta^{3}y^{2}}\right)}.
\ee}
The first integral corresponding to (\ref{actionLC}) is
\be
\mathcal{H} = \frac{2a^2 M\cos^4\Theta-\Delta^{3} y^{2}(a^2+(a^2-\sin^{2}\Theta)y^{2})}{4a^{2}\Delta^{3}\mathcal{L}}\cos^2\Theta =- C,
\ee
where $\mathcal{L}$ is the Lagrangian corresponding  to the action (\ref{actionLC}).
Then the equation of motion is given by
{\small
\be\label{yprime}
y'^{2} = \frac{(2M - \Delta^{2}y^{2}(1+y^2))(2a^2(2C^2\Delta^3 +2M \cos^{6}\Theta)- \Delta^3\cos^{2}\Theta y^{2}(a^{2} + y^{2}(a^{2}-\sin^{2}\Theta)))}{4a^{2}C^{2}\Delta^{5}}\cos^{2}\Theta,
\ee}
where $y' = \frac{\partial y}{\partial \sigma^{1}}$ and $C$  is the constant energy of motion.

Plugging  (\ref{yprime}) back into (\ref{actionLC}) yields to
\be\label{actionPsi}
S =\frac{L^{-}}{2\pi\alpha'}\int^{ L/2}_{-L/2}d\Psi \frac{\cos^2\Theta }{4C}  \left(y^2 (1 +  y^2)-\frac{y^4 \sin ^2\Theta }{a^2}-\frac{2 M \cos ^4\Theta }{\Delta ^3}\right).
\ee

Remembering that $ d\Psi =\displaystyle{ \frac{dy}{y'}}$
we come to the integration in terms  of $y$, so we get for (\ref{actionPsi})
{\footnotesize
\be\label{actNGdy}
S=\frac{L^{-}}{\pi\alpha'}\int^{\infty}_{y_{H}}dy\frac{a \Delta ^{5/2} \cos \Theta y^2 \left(1+ y^2 -\frac{y^2 \sin ^2\Theta}{a^2}-\frac{2 M \cos ^4\Theta }{\Delta ^3 y^2}\right)}{2\sqrt{(2M -\Delta^{2}y^{2}(1+y^2))}\sqrt{2a^2(2C^2\Delta^3 +M\cos^{6}\Theta)-\Delta^3\cos^{2}\Theta y^{2}(a^{2}(1 + y^{2})-y^2\sin^{2}\Theta))}} .
\ee}


The action (\ref{actNGdy}) contains divergences on the spacetime boundary $y\to +\infty$. To renormalize this we have to subtract from (\ref{actNGdy}) the mass of two single quarks, given by the term
 \be\label{S0finT}
 S_{0} =  \frac{L^{-}}{\pi\alpha'} \int^{\infty}_{y_{H}}dy \sqrt{|G_{x^{-}x^{-}}G_{yy}|} = \frac{L^{-}}{2\pi\alpha'} \int^{\infty}_{y_{H}}dy \sqrt{\frac{1+y^{2}- \frac{y^{2}\sin^{2}\Theta}{a^{2}}-\frac{2M}{\Delta^{3}y^{2}}\cos^{4}\Theta}{1+y^{2} - \frac{2M}{\Delta^{2}y^{2}}}}.
 \ee
Subtracting  (\ref{S0finT}) from (\ref{actNGdy}) we get the regularized Nambu-Goto action
\bea\label{Sreg}
&& S - S_{0} = \frac{L^{-}}{\pi\alpha'}\int^{\infty}_{y_{H}}dy \sqrt{\frac{1+ y^2-\frac{y^2 \sin ^2\Theta}{a^2}-\frac{2 M \cos ^4\Theta }{\Delta ^3 y^2}}{1+y^{2} - \frac{2M}{\Delta^{2}y^{2}}}}
 \times\\
 &&
 \left(\frac{a \Delta^{3/2} \cos \Theta y \sqrt{1+ y^2-\frac{y^2 \sin ^2\Theta}{a^2}-\frac{2 M \cos ^4\Theta }{\Delta ^3 y^2}}}{2\sqrt{\Delta^3\cos^{2}\Theta y^{2}(a^{2}(1 + y^{2})-y^2\sin^{2}\Theta))-2a^2(2C^2\Delta^3 +M\cos^{6}\Theta)}}  - 1\right) .\nonumber
\eea
Expanding in series assuming that  $M$ is large we get at first order
\be\label{Sregexp}
S^{reg}=S -S_{0}\approx - \frac{L^{-}}{\pi\alpha'} \int^{\infty}_{y_{H}}dy \frac{C^2 \Delta^{5/2}\sec^{4}\Theta}{2M} + \mathcal{O}(M^2).
\ee
  Integrating  (\ref{Sregexp}) in large  $M$ we get
 \be
 S^{reg}\approx \sqrt{\lambda}\frac{L^{-}}{\pi}\frac{C^2 \Delta^{5/2}\sec^{4}\Theta}{2M}y_{+}+ \mathcal{O}(M^2),
 \ee
  where we define $\lambda = \frac{1}{\alpha'^{2}}$. For small $a$ it has a simple form
  \be\label{SS0}
  S^{reg}\approx \sqrt{\lambda}\frac{L^{-}}{\pi}\frac{C^2 \sec^{4}\Theta}{2M}y_{H} + \mathcal{O}(a^2).
  \ee 
To derive $C$ we find the following relation for the length $L$ from (\ref{yprime})
 {\footnotesize
\be\label{length2}
\frac{L}{2}= \int_{y_{H}}^{\infty}dy \frac{2aC\Delta^{5/2}}{\sqrt{( \Delta^{2}y^{2}(1+y^2)-2M)(-2a^2(2C^2\Delta^3 +2M \cos^{6}\Theta)+\Delta^3\cos^{2}\Theta y^{2}(a^{2} + y^{2}(a^{2}-\sin^{2}\Theta)))}\cos\Theta}.
\ee}
at hight temperature limit ($M\to \infty$) eq. (\ref{length2}) takes the form 
\be\label{lengthC}
\frac{L}{2}\approx - \int_{y_{H}}^{\infty}dy \frac{C\Delta^{5/2}\sec^{4}\Theta}{M} + \mathcal{O}(M^2)  \approx  \frac{C\Delta^{5/2}\sec^{4}\Theta}{M}y_{H}+ \mathcal{O}(M^2),
\ee
or for small $a$
\be\label{constCSSI}
\frac{ L}{2C} = -\frac{\sec^{4}\Theta}{M}y_{H} + \mathcal{O}(a^2).
\ee
Then from (\ref{SS0}) and (\ref{constCSSI}) we find that the regularized action is
\be\label{2SI0}
2S_{I}^{reg} =  \frac{\sqrt{\lambda}}{\pi}\frac{L^{-}L^2 M\cos^4\Theta}{4y_{H}}.
\ee
Taking into account that $M = \frac{y^2_{H}(y^2_{H} + 1)}{2}$ in terms of $T_{H}$, related to $y_{H}$ as $y_{H} = (\pi T_{H} + \sqrt{\pi^2 T^{2}_{H}- 2})/2$ we have
\begin{align}\label{2SI0regnew}
2S_{I}^{reg} = \frac{\sqrt{\lambda }}{32\pi }L^{-}L^{2} (\sqrt{\pi ^{2} T^{2}_{H}-2}+
\pi T_{H}) (\pi T_{H} (\sqrt{\pi ^{2} T^{2}_{H}-2}+\pi T_{H})+1)
\cos ^{4}\Theta .\
\end{align}
By virtue of (\ref{JQPW}), (\ref{JQPS}) and (\ref{2SI0regnew}) we find that the
jet-quenching parameter is given
\be
\hat{q} =\sqrt{\lambda}\frac{\left(\sqrt{\pi ^2 T^2_{H}-2}+\pi  T_{H}\right) \left(\pi  T_{H} \left(\sqrt{\pi ^2 T^2_{H}-2}+\pi  T_{H}\right)+1\right)
   \cos^4\Theta }{4 \sqrt{2}\pi},
\ee
which at high temperature takes the form
\be\label{jetQP2}
\hat{q} =
 \frac{\sqrt{\lambda}}{\sqrt{2}} \pi ^2  T^3_{H} \cos^4\Theta.
\ee

In (\ref{jetQP2}) we observe that the jet-quenching parameter depends on the temperature as $T^{\,\,3}_{H}$ 
and linearly on $\sqrt{\lambda}$ as in the case of the planar black brane \cite{HoloJet}.

\section{Summary and Discussion}

In the present paper, we have investigated the energy loss of a heavy quark in a rotating quark-gluon plasma within the holographic framework.
As a holographic dual we have considered the 5d Kerr-AdS black hole with two arbitrary rotational parameters.  The Kerr-AdS black hole is justified for the description of the QGP at  high temperature, where it is believed that the conformal symmetry is restored. 
Using the holographic prescription we have calculated  the energy of a static quark in both the rotating- and static-at-infinity frames under the assumption that the rotational parameters are small. These relations gereneralize our result \cite{AGG2020} obtained for a single rotational parameter (in the BL coordinates). We show that for small values of the rotational parameters the mass of a static quark in the rotating- and static-at-infinity frames, are similar.
 We have shown that at high temperature the contribution from the rotation in both coordinate systems is suppressed,  thus the dependence of the energy on the temperature turns to be the linear relation $\frac{\sqrt{\lambda}}{2} T_H$ like for the case of the planar black brane  \cite{HKKKY}. It is interesting to note that in BL coordinates (the rotating-at-infinity frame), for zero temperature the Lagrangian mass of the static quark has a contribution from the rotational parameters, while in non-rotating-at-infinity frame (AdS coordinates) it does not.

 We have also presented the general formulae for the holographic drag forces acting on a heavy quark for the case of two arbitrary rotational parameters. 
In the holographic description  the heavy quark can  be considered through a string  hanging towards the black hole horizon with a fixed endpoint attached to the boundary of the Kerr-${\rm AdS}_5$ background. We have considered the rotating-at-infinity frame of the Kerr-${\rm AdS}_{5}$ with the string endpoint  fixed on the boundary, so the string rotates together with the QGP and the static-at-infinity frame with a slowly moving endpoint of the string. We have shown that the results for the drag force are equivalent in both frames and have a linear dependence on the temperature $T_{H}$. The only difference is a numerical factor, which comes from the fact that we have used the approximated form of the metric for the static-at-infinity frame. We have seen that the drag force has a contribution from the difference of squares of  the angular velocities as in the hydrodynamical calculations for the pressure gradient \cite{AGG2020}. The drag force relations are also in agreement with the particular cases, i.e. a single non-zero rotational parameter and two equal rotational parameters,  obtained in the previous paper \cite{AGG2020}. For the case of two equal  rotational parameters, which includes the extremal case,  we have found that the drag force vanishes. This was also observed in  \cite{AGG2020} using the hydrodynamical calculations.
Note that for the case of a single rotational parameter in the rotating-at-infinity frame, the relation for the drag force matches with that one from \cite{NAS}.

In the case of the static-at-infinity frame and a single rotational parameter $a$, we have estimated the jet-quenching parameter and found that the leading term (for a small $a$) has the same dependence on $\sqrt{\lambda}$ and $T_{H}$ as in the case of a planar AdS black brane \cite{HoloJet}. 

As an extension for future work,
it will be interesting to study the jet-quenching parameter more carefully  in both the rotating- and static-at-infinity frames, to compute and analyze the contributions from the rotation of a higher order.

Another natural direction is to generalize the results for the energy losses on the case of  the rotating charged
AdS black holes~\cite{Cvetic}, which correspond to a conformal plasma with non-zero chemical potential. It would be important to trace and analyze the influence of the chemical potential and the second non-zero rotational parameters on the transport properties of the QGP.  It will be also of interest to compare results with those for the rotating quark in $\mathcal{N}=4$ SYM (static) quark-gluon plasma at finite chemical potential from \cite{Hou:2021own}. 
 
\section*{Acknowledgments}
We are grateful to I. Ya. Aref'eva, V. Braguta, H. Dimov, Yu. Ivanov and  O. Teryaev for useful stimulating discussions and comments.
This work (A.G.) is supported by RFBR Grant 18-02-40069 (mega). A.G is also grateful to the Paris Observatory, LUTh for kind hospitality, where a part of this work was done.

\appendix

\newpage
\setcounter{equation}{0} \renewcommand{\theequation}{A.\arabic{equation}}
\section{Metric components}
\subsection{Boyer-Lindquist coordinates}
The components of the metric (\ref{2}) in general are
\bea\label{3}
G_{tt}&=& \frac{1}{\rho^2} \left(-\Delta_r + \Delta_{\theta}\left(a^2\sin^2\theta+b^2\cos^2\theta\right)+\frac{a^2b^2(1+r^2\ell^2)}{r^2}\right),\\
G_{rr}&=& \frac{\rho^{2}}{\Delta_r},\, \quad G_{\theta\theta}=\frac{\rho^{2}}{\Delta_{\theta} },\\
G_{\phi\phi}&=&\frac{\sin^{2}\theta}{\rho^2\Xi^{2}_a}\left(-a^2\sin^2\theta\Delta_r+(r^2+a^2)^2(\Delta_{\theta}+\frac{b^2\sin^2\theta}{r^2}(1+r^2\ell^2))\right),\\
G_{\psi\psi}&=&\frac{\cos^{2}\theta}{\rho^2\Xi^{2}_b}\left(-b^2\cos^2\theta\Delta_r+(r^2+b^2)^2(\Delta_{\theta}+\frac{a^2\cos^2\theta}{r^2}(1+r^2\ell^2))\right),\\
G_{t\phi}&=&G_{\phi t}=\frac{a\sin^{2}\theta}{\rho^2\Xi_a}\left(\Delta_r-(r^2+a^2)(\Delta_{\theta}+\frac{b^2}{r^2}(1+r^2\ell^2))\right),\\
G_{t\psi}&=&G_{\psi t}=\frac{b \cos^{2}\theta}{\rho^2\Xi_b}\left(\Delta_r-(r^2+b^2)(\Delta_{\theta}+\frac{a^2}{r^2}(1+r^2\ell^2))\right),\\ \label{3phipsi}
G_{\phi\psi}&=&G_{\psi\phi}=\frac{abM\sin^{2}(2\theta)}{2\rho^{2}\Xi_a\Xi_b}.
\eea

If we suppose that $b=an$ then the metric components (\ref{3})-(\ref{3phipsi})  are transformed to
\bea\label{metriccomp}
G_{\phi\phi}&=&\frac{\sin^{2}\theta}{\rho^2\Xi^{2}_a}\left(-a^2\sin^2\theta\Delta_r+(r^2+a^2)^2(1 - a^2\ell^2 \cos^{2}\theta + \frac{1}{r^{2}}n^2 a^2\sin^2\theta )\right),\\
G_{\psi\psi} &=&\frac{\cos^{2}\theta}{\rho^2\Xi^{2}_b}\left(-n^2 a^2\cos^2\theta\Delta_r+(r^2+ n^2a^2)^2(1 - a^2n^2\ell^2 \sin^{2}\theta +\frac{1}{r^2}a^2\cos^2\theta)\right)\\
G_{t\phi}&=&G_{\phi t}=\frac{a\sin^{2}\theta}{\rho^2\Xi_a}\left(\Delta_r-(r^2+a^2)(1 + \frac{n^2a^2}{r^{2}} -a^2\ell^2\cos^{2}\theta(1-n^2))\right),\\
G_{t\psi} &=&G_{\psi t}=\frac{n a \cos^{2}\theta}{\rho^2\Xi_b}\left(\Delta_r-(r^2+n^2 a^2)(1+\frac{a^2}{r^2}+ a^2\ell^2 \sin^2 \theta (1-n^2) )\right)\\
G_{\phi\psi}&=&G_{\psi\phi}=\frac{na^2 M\sin^{2}(2\theta)}{2\rho^{2}\Xi_a\Xi_b},\quad G_{rr}= \frac{\rho^{2}}{\Delta_r},\quad G_{\theta\theta}=\frac{\rho^{2}}{\Delta_{\theta} },\\ \label{metriccomp2}
G_{tt}&=& \frac{1}{\rho^2} \left(-\Delta_r + a^2\Delta_{\theta}\left(\sin^2\theta+n^2\cos^2\theta\right)+\frac{n^2 a^4(1+r^2l^2)}{r^2}\right).
\eea

\subsection{Asymptotycally AdS coordinates}

The components of the metric (\ref{metlim}):
\bea\label{compAdS}
G_{TT}&=& \frac{2M}{\Delta^{3}y^{2}}-(1+y^{2}),\quad G_{yy}= \frac{1}{1+y^{2} - \frac{2M}{\Delta^{2}y^{2}}} ,\quad G_{\Theta\Theta}=y^{2},\\
G_{\Phi\Phi}&=& y^{2}\sin^{2}\Theta+\frac{2M a^{2}\sin^{4}\Theta}{\Delta^{3}y^{2}},\quad  G_{\Psi\Psi}=y^{2}\cos^{2}\Theta+\frac{2Mb^{2}\cos^{4}\Theta}{\Delta^{3}y^{2}},\\
G_{T\Phi}&=&-\frac{2Ma\sin^{2}\Theta}{\Delta^{3}y^{2}},\quad  G_{T\Psi}=- \frac{2Mb\cos^{2}\Theta}{\Delta^{3}y^{2}},\quad G_{\Phi\Psi}=\frac{2Mab\sin^{2}\Theta\cos^{2}\Theta}{\Delta^{3}y^{2}}.
\eea
\subsection{Light-cone coordinates}
 The metric components in light-cone coordinates are given by
\bea\label{metricLC}
G_{x^{-}x^{-}}&=&-\frac{1}{4}\Big(1+y^{2}- \frac{y^{2}\sin^{2}\Theta}{a^{2}}-\frac{2M}{\Delta^{3}y^{2}}\cos^{4}\Theta\Bigr),\\
G_{x^{+}x^{+}}&=&-\frac{1}{4}\Bigl(1+y^{2} -\frac{ y^{2}\sin^{2}\Theta}{a^{2}} -\frac{2M}{\Delta^{3}y^{2}}(1+\sin^{2}\Theta)^{2}\Bigr) ,\\
G_{x^{+}x^{-}}&=&-\frac{1}{4}\left(1+y^{2}+\frac{y^{2}\sin^2\Theta}{a^2} - \frac{2M}{\Delta^{3}y^{2}}(1-\sin^{4}\Theta )\right)\\
G_{yy}&=& \frac{1}{1+y^{2} - \frac{2M}{\Delta^{2}y^{2}}} ,\quad G_{\Theta\Theta}=y^{2},\quad G_{\Psi\Psi}=y^{2}\cos^{2}\Theta.
\eea

\section{The limits of 5d Kerr-AdS metric. Special cases}
\setcounter{equation}{0} \renewcommand{\theequation}{B.\arabic{equation}}

For zero parameters $a$ and $b$ the metric \ref{2} is simplified:
\bea\label{SchwAdS2}
ds^2&=& -r^2f(r){dt}^2+r^2\left(d\theta^2+\frac{{dr}^2 }{r^4 f(r)}+ \cos ^2\theta d\psi^2+ \sin^2\theta d\phi^2\right),
\eea
where $f(r)= l^2+\frac{1}{r^2}-\frac{2M}{r^4}$ is the blackening function. This is nothing but the 5d Schwarzchild-AdS metric with the horizon located at
\be\label{SchwAdShor}
r_{H} = \frac{\sqrt{\sqrt{1 +8\ell^2 M }-1}}{\sqrt{2}\ell},
\ee
which is calculated as the largest root of $f(r) =0$.  So for the solution (\ref{SchwAdS2}) the radial coordinate runs as $r\in(r_{H},+\infty)$.
The Hawking temperature is given by
\be\label{rhschwT}
 T_{H} = \frac{4 r^2_{H}\ell^2 + 2  }{4\pi r_{H}},
\ee
 the location of the horizon is defined through the temperature as
\be\label{rhschwTR}
r_{H} = \frac{1}{2\ell^2}\left(\pi T_{H} + \sqrt{\pi^2 T^{2}_{H}- 2\ell^2}\right).
\ee
From (\ref{SchwAdShor}) one can write down some useful relations
 \bea\label{usrel4Mlm2}
 && 2\ell^2 r^{2}_{H}  = \sqrt{1+8\ell^2 M} -1,\quad 2\ell^2 r^{2}_{H}+ 1 = \sqrt{1+8\ell^2 M},\\  \label{usrel4Ml3}
&& 1+4\ell^2 M - \sqrt{1+8\ell^2 M} = 2\ell^4 r^4_H, \quad 4(\ell^2 r^{2}_{H}+ 1)^2 = (\sqrt{1+8\ell^2 M} + 1)^2.
\eea
We note that $r_{H} =0$ if
\be
M =0,
\ee
in this case the metric (\ref{2}) is just the 5d AdS spacetime in the global coordinates.

It is also instructive to show that the metric (\ref{metlim}) with $a= b=0$ has $\Delta = 1$ comes to the following form
	\begin{equation}\label{metlimC}
ds^{2}\simeq -(1+y^{2}-\frac{2M}{y^{2}})dT^{2} + \frac{dy^{2}}{1+y^{2} - \frac{2M}{y^{2}}} + y^{2}(d\Theta^{2} + \sin^{2}\Theta d\Phi^{2} + \cos^{2}\Theta d\Psi^{2}).
	\end{equation}

\end{document}